\documentclass[aps,prl,twocolumn,groupedaddress,amsmath,amssymb]{revtex4-1}
\usepackage{graphicx}  
\usepackage{dcolumn}   
\usepackage{bm}        
\usepackage{verbatim}   
\usepackage{bbold}
\usepackage{color}
\begin{document}

\title{Optical Conductivity From Pair Density Waves}
\author{Zhehao Dai}
\affiliation{
   Department of Physics,
   Massachusetts Institute of Technology,
   Cambridge, MA 02139, USA
   }
\author{Patrick A. Lee}
\affiliation{
   Department of Physics,
   Massachusetts Institute of Technology,
   Cambridge, MA 02139, USA
   }

\date{\today}
\begin{abstract}
We present a theory of optical conductivity in systems with finite-momentum Cooper pairs. In contrast to the BCS pairing where AC conductivity is purely imaginary in the clean limit, there is nonzero AC absorption across the superconducting gap for finite-momentum pairing if we break the Galilean symmetry explicitly in the electronic Hamiltonian. Vertex correction is crucial for maintaining the gauge invariance in the mean-field formalism and dramatically changes the optical conductivity in the direction of the pairing momentum.  We carried out a self-consistent calculation and gave an explicit formula for optical conductivity in a simple case. This result applies to the Fulde-Ferrell-Larkin-Ovchinnikov state and candidates with pair density waves proposed for High-Tc cuprates. It may help detect PDW and determine the pairing gap as well as the direction of the pairing momentum in experiments.
\end{abstract}

\maketitle

\section{Introduction} \label{introduction}
Pair density waves (PDW) occur when Cooper pairs condense at nonzero momenta. The first example of PDW is the Fulde-Ferrell-Larkin-Ovchinnikov state (FFLO), where finite-momentum pairing is preferred in a certain range of the Zeeman splitting \cite{fulde1964superconductivity,larkin1965inhomogeneous}. More recently, experimental evidence of FFLO states has been found in $\text{CeCoIn}_{5}$ \cite{bianchi2003possible} and $\kappa\text{-(BEDT-TTF)}_{2}\text{Cu(NCS)}_{2}$ \cite{mayaffre2014evidence}, and possible mechanisms stabilizing PDW have been proposed in High-Tc cuprates \cite{berg2009striped,PhysRevX.4.031017}. Unlike conventional BCS superconductors, these phases with PDW usually have partially-gapped Fermi surfaces, almost normal specific heat and anisotropic electromagnetic response. Although many of the physical properties of PDW are well-established, to the best of our knowledge, the optical conductivity from PDW have not yet been addressed. The purpose of the present paper is to report the unconventional features in the optical conductivity and to discuss its potential applications in various experimental systems. Most of the results presented here apply to a general class of PDW, but we mainly focus on the case with FFLO pairing where quantitative comparison might be made with experiments in the near future.

It is well-known that a single-band BCS superconductor, in the clean limit, has no optical absorption across the superconducting gap \cite{mahan2013many}. This absence of absorption is not protected by the symmetry of the Hamiltonian but by a special feature of the BCS ground state: single-particle states in the original band carrying opposite currents are always simultaneously occupied (or unoccupied), hence the ground state is an exact eigenstate of the current operator and the matrix element for AC absorption $\langle \text{excited state} |\ \vec{\mathbf{j}}\ |\text{G.S.}\rangle$ (often called the `coherence factor') vanishes.
However, this is not the case for finite-momentum pairing. Although the ground state has zero average current, it is no longer an eigenstate of the current operator. Finite-momentum Cooper pairs are in general optically active and they give rise to the dominant contribution to the AC conductivity in the energy range comparable to the pairing gap.

It is worth mentioning that the ground state generally involve PDW with multiple pairing momenta if finite-momentum pairing is favorable. For example, if we have Cooper pairs condensing at momentum $Q$, it's natural to have another pairing momentum $-Q$. The two pairing terms together cause the folding of the Brillouin zone (B.Z.), hence charge density waves (CDW) at momenta $2Q$, $4Q$ etc  \cite{larkin1965inhomogeneous}. It is also possible to have pairing momenta in different directions generating complex incommensurate patterns above the original lattice. However, for simplicity, we focus on the case with only one pairing momentum, a `pure PDW' with no charge modulation. The optical absorption from PDW with multiple pairing momenta should be qualitatively similar for frequencies around the pairing gap. This `pure PDW' with only a phase modulation in the pairing order parameter appears to break the lattice translational symmetry, but it is actually invariant under the combination of a gauge transformation and the lattice translation. Note that the absolute phase is not a physical observable, only the phase difference is. Despite the phase modulation, every physical observable in this state is invariant under the lattice translation. In this sense, we do not need to break the translational symmetry further to get new absorption peaks, this is very different from the optical absorption of CDW only.

One important thing in calculating optical conductivity is maintaining gauge invariance in the self-consistent main-field approximation. This issue was first discussed in BCS superconductors by Nambu \cite{nambu1960quasi}, and recently studied in strongly interacting superconductors \cite{he2015establishing,boyack2016gauge}. The key step is to carry out the vertex correction that is consistent with the gap equation \cite{nambu1960quasi,schrieffer1964theory,he2015establishing,boyack2016gauge}. We followed Nambu's approach and gave an explicit formula for optical conductivity in systems with simple electron-electron interactions. One subtlety in this calculation is that, in order to have non-zero AC conductivity, we must break Galilean symmetry explicitly in the electronic Hamiltonian. This issue is discussed in more detail after a brief review on finite-momentum pairing.

\section{Finite-momentum pairing and gap equation}

We start by briefly reviewing the mean-field treatment of finite-momentum pairing, especially the diagrammatic interpretation of the mean field gap equation, which turned out to be useful in calculating linear response functions.

\begin{figure}
\begin{center}
\includegraphics[width=3in]{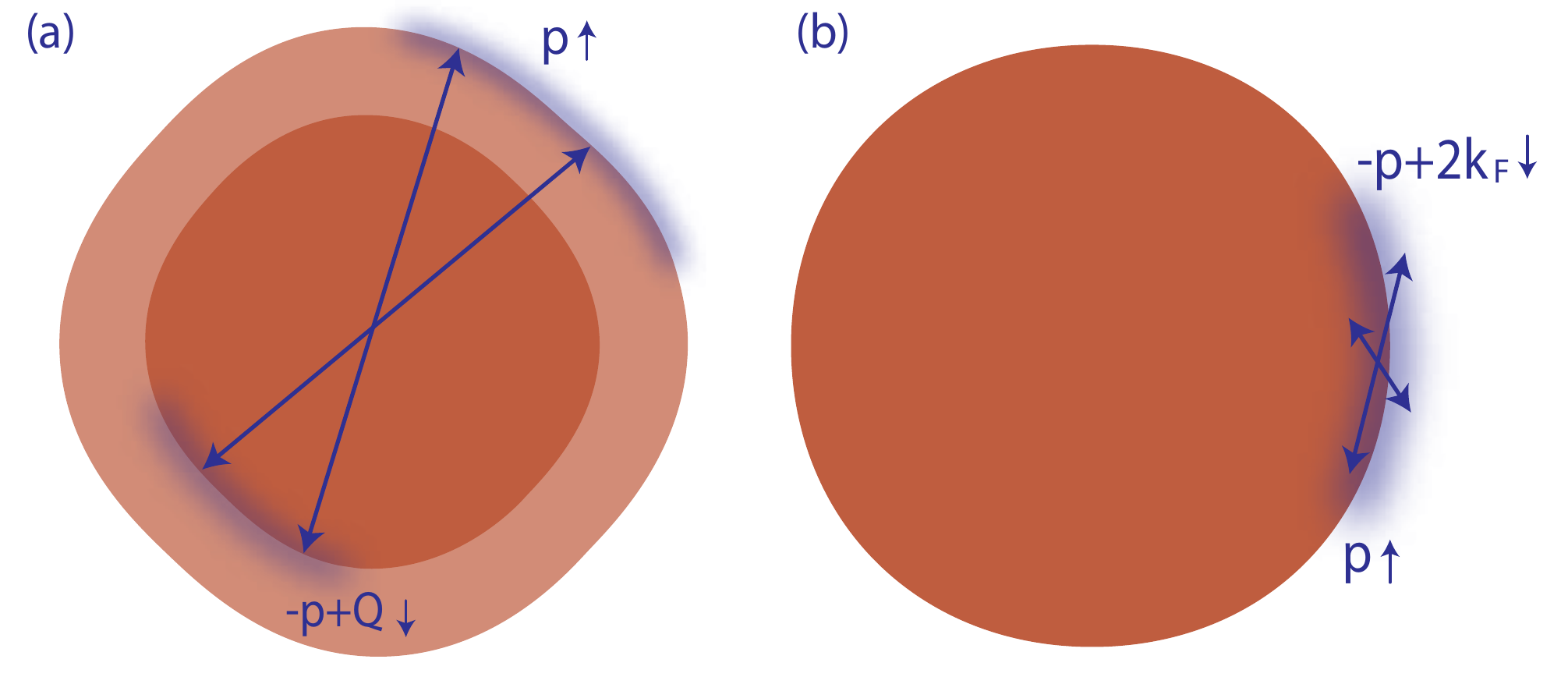}
\caption{Two examples of finite-momentum pairing. (a) FFLO pairing. The dark orange region is occupied by both spins, while the light orange region is occupied by up spin only. The blue shaded regions on the Fermi surface are gapped out by pairing. (b) Amperean pairing. A different pairing mechanism without spin-splitting, where the vicinity of a hot spot on the Fermi surface is gapped out, and the pairing momentum is close to $2k_{F}$.}
\label{fig:pairing}
\end{center}
\end{figure}

In the case of FFLO pairing, the Fermi surfaces of up-spin and down-spin electrons are split by Zeeman splitting, but the orbital degree of freedom is not affected. This situation can be realized in layered materials by imposing an in-plane magnetic field. As shown in figure \ref{fig:pairing}(a), finite-momentum pairing creates Cooper pairs near the Fermi surfaces, and is argued to be more stable than the BCS pairing in a certain range of spin-splitting. Another example of finite-momentum pairing is the Amperean pairing shown in figure \ref{fig:pairing}(b), where electrons moving in the same direction attract each other by the Lorentz force of the emergent gauge field \cite{lee2007amperean,PhysRevX.4.031017}. 

In the present paper, We consider a (2+1)-dimensional system with Hamiltonian $H=\sum\epsilon_{p,\sigma}\psi^{\dagger}_{p,\sigma}\psi_{p,\sigma} + \sum\lambda_{k}\psi_{p+k,\sigma}^{\dagger}\psi_{p'-k,\sigma'}^{\dagger}\psi_{p',\sigma'}\psi_{p,\sigma}$, where the four-Fermion interaction might be mediated by phonon or other more exotic mechanisms. To describe a state with finite-momentum pairing, it is convenient to introduce the 2-component Nambu spinor: $\Psi_{p}=(\psi_{p+Q/2,\uparrow},\ \psi_{-p+Q/2,\downarrow}^{\dagger})^{T}$, where $Q$ is the paring momentum which should be determined self-consistently. The four-Fermion interaction can then be written as
$\sum_{p,p',k}\lambda_{k}[\Psi^{\dagger}_{p+k}\tau_{3}\Psi_{p}][\Psi^{\dagger}_{p'-k}\tau_{3}\Psi_{p'}]$.
The mean field Hamiltonian for finite-momentum pairing is:
\begin{equation}
\label{eq:mean field Hamiltonian}
H=\sum_{p}\Psi_{p}^{\dagger}\left(\begin{array}{cc}
\epsilon_{p+Q/2,\uparrow} & \Delta_{p}\\
\Delta_{p} & -\epsilon_{-p+Q/2,\downarrow}
\end{array}\right)\Psi_{p}
\end{equation}
We would like to point out an important difference with the BCS pairing. In the BCS case, the diagonal terms are always equal with opposite signs, so are the two eigenvalues. However, this `particle-hole' symmetry is broken in the FFLO state. We may even have an `unpaired region' in the B.Z. where the two eigenvalues are of the same sign. For convenience, define $\bar{\epsilon}_{p}\equiv(\epsilon_{p+Q/2,\uparrow} + \epsilon_{-p+Q/2,\downarrow})/2$, $\epsilon'_{p}\equiv(\epsilon_{p+Q/2,\uparrow} - \epsilon_{-p+Q/2,\downarrow})/2$, and $\delta_{p}\equiv\sqrt{\bar{\epsilon}_{p}^{2}+\Delta_{p}^{2}}$. The two eigenvalues are given by
\begin{equation}
E_{p}^{\pm}=\epsilon'_{p}\pm\delta_{p}
\end{equation}
The unpaired region is where $\delta_{p}<|\epsilon'_{p}|$. The boundary of this region where $\delta_{p}=|\epsilon'_{p}|$ is the `Fermi Surface' left after FFLO pairing and the shifting in momentum. Optical absorption occurs in the `paired region' when the frequency of light matches the splitting between the two bands $2\delta_{p}$.

The Nambu spinor introduced above allows us to treat the pairing gap on an equal-footing with the self-energy correction, and the conventional mean field gap equation can be understood as a Hatree-Fock approximation \cite{nambu1960quasi,schrieffer1964theory}. We approximate the four-Fermion interaction by a quadratic term and demand that, to the first order, the remaining interaction does not modify the propagator:
\begin{eqnarray}
\left\{\begin{array}{ll}
G(p)=1/(p_{0}-H_{0}(p)-\Sigma(p) + isgn(p_{0})0^{+})\\
0=-\Sigma(p) + i\int\frac{d^{3}k}{(2\pi)^3}\lambda_{k}\tau_{3}G(p-k)\tau_{3}
\end{array}\right.
\label{eq:mean field Green}
\end{eqnarray}
where $G(p)$ is the mean-field Green's function of the Nambu spinor, $p_{0}$ is the temporal component of the momentum, $H_{0}(p)\equiv \epsilon'_{p} + \bar{\epsilon}_{p}\tau_{3}$ is the Hamiltonian for the original band, and $\Sigma(p)\equiv\Delta_{p}\tau_{1}$ is the pairing term. We have ignored the diagonal self-energy correction in $\Sigma(p)$ since it is not important for our purpose.

\begin{figure}
\begin{center}
\includegraphics[width=3in]{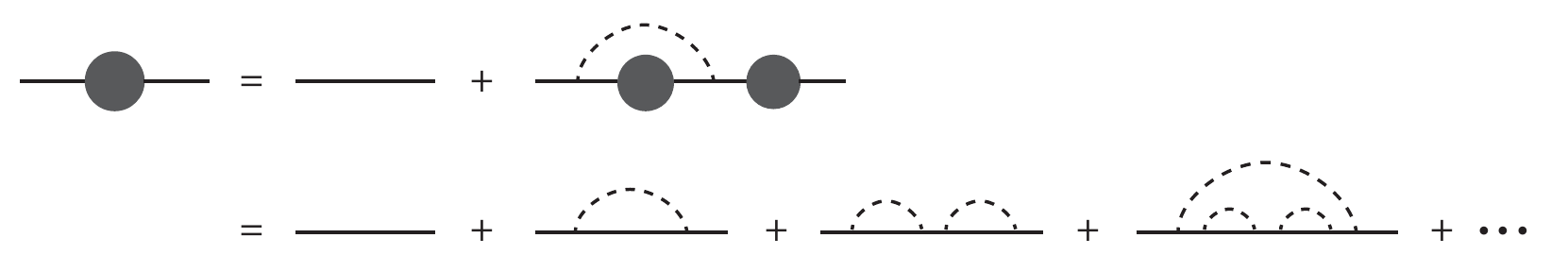}
\caption{The self-consistent equation of the mean field Green's function, and the diagrams included in this approximation. The solid line represents the 2-component Nambu spinor, and the dashed line represents the electron-electron interaction mediated by a boson, e.g. phonon. We have ignored the correction of the interaction, since it is not important for our purpose. All diagrams without the crossing of the interaction line is included.}%
\label{fig:selfenergy}
\end{center}
\end{figure}

This approximation is equivalent to summing over all Feynman diagrams without crossing in calculating the Green's function, as shown in figure \ref{fig:selfenergy}.

When the four-Fermion interaction has no momentum dependence near the Fermi surface, both $\lambda_{k}$ and $\Delta_{p}$ can be approximated by constants, and we arrive at the familiar gap equation after integrating out $k_{0}$:
\begin{equation}
\Delta=-\lambda\int_{\text{paired}}\frac{d^{2}\vec{\mathbf{p}}}{(2\pi)^{2}}\frac{\Delta}{2\sqrt{\bar{\epsilon}_{p}^{2} + \Delta^{2}}}
\end{equation}
This gap equation is almost the same as the BCS gap equation, except the integral is restricted in the `paired region'.

\section{Vertex correction and gauge invariant electromagnetic response}

We are now ready to study the electromagnetic response of PDW. Following the Peierls substitution, we change $\epsilon_{p,\sigma}$ in the total Hamiltonian into $\epsilon_{p+eA,\sigma}$, where $\vec{\mathbf{A}}$ is the magnetic vector potential. We restrict ourself to the single band near the Fermi level, and focus on the limit of a weak and uniform external field as in the case of infrared absorption. Under these restrictions, the current operator $\vec{\mathbf{j}}\equiv-\partial H/\partial\vec{\mathbf{A}}$ can be written as
\begin{eqnarray}
\vec{\mathbf{j}}&=&\sum_{p,\sigma}\psi_{p,\sigma}^{\dagger}[-e\vec{\mathbf{v}}_{p,\sigma}-e^{2}\mathbf{m}_{p}^{-1}\vec{\mathbf{A}}]\psi_{p,\sigma}\label{eq:current}\\
&\equiv&\sum_{p}\Psi_{p}^{\dagger}[-e\vec{\mathbf{v}}_{1}(\vec{\mathbf{p}})\mathbb{1} - e\vec{\mathbf{v}}_{2}(\vec{\mathbf{p}})\tau_{3} -e^{2}\mathbb{m}_{p}^{-1}\vec{\mathbf{A}}]\Psi_{p}\label{eq:current,Nambu spinor}
\end{eqnarray}
where $\vec{\mathbf{v}}_{p,\sigma}\equiv\nabla_{p}\epsilon_{p,\sigma}$ is the band velocity and $\mathbf{m}_{p}\equiv(\nabla_{p}\nabla_{p}\epsilon_{p,\sigma})^{-1}$ is the effective mass tensor. $\vec{\mathbf{v}}_{1}(\vec{\mathbf{p}})$, $\vec{\mathbf{v}}_{2}(\vec{\mathbf{p}})$ and $\mathbb{m}_{p}$ are defined by the equation above and they depend on the pairing momentum. The current operator at zero field is usually called the paramagnetic current, and we would like to write the spatial components together with the temporal component $j_{0}=\sum_{p,\sigma}-e\psi_{p,\sigma}^{\dagger}\psi_{p,\sigma}$ as:
\begin{eqnarray}
j^{P}_{\mu}&=&\sum_{p}\Psi_{p}^{\dagger}\gamma_{\mu}(\vec{\mathbf{p}})\Psi_{p},\label{eq:paramagnetic current}\\
\gamma_{\mu}(\vec{\mathbf{p}})&\equiv&-e(\tau_{3},\ \vec{\mathbf{v}}_{1}(\vec{\mathbf{p}})\mathbb{1} + \vec{\mathbf{v}}_{2}(\vec{\mathbf{p}})\tau_{3})\label{eq:bare vertex}
\end{eqnarray}
The part of current proportional to $\vec{\mathbf{A}}$ in equation \ref{eq:current} and \ref{eq:current,Nambu spinor} is called the diamagnetic current, which does not contribute to the real part of the conductivity at any finite frequency.

Naively, one would like to plug the paramagnetic current and the mean-field excited states into the Kubo formula:
\begin{equation}
\label{eq: bare result}
\text{Re}\,\sigma_{ii}=\frac{\pi}{\omega}\sum_{n}|\langle 0|j^{P}_{i}|n\rangle|^2\delta(\omega-E_{n}+E_{0})
\end{equation}
where $i$ denotes the spatial components, and $0$ ($n$) denotes the ground state (excited states). This approach corresponds to plugging the mean-field Green's function into the bubble diagram without doing other corrections.

As explained in the introduction, the matrix element $\langle 0|j^{P}_{i}|n\rangle$ vanishes identically for BCS pairing, but not for finite-momentum pairing. Thus we expect a nonzero AC conductivity for a state with PDW. However the bare result given by the `mean-field-version' of equation \ref{eq: bare result} can not be trusted for at least two reasons: (1) This approach violates gauge invariance, specifically the Ward-Takahashi identity between the vertex and the Green's function \cite{nambu1960quasi,schrieffer1964theory}. (2) The result given by equation \ref{eq: bare result} is always nonzero for any finite-momentum pairing, but the AC conductivity should be exactly zero if the electronic Hamiltonian is Galilean invariant.

The latter statement may not be immediately obvious, especially in the case with spontaneous symmetry-breaking. So we give a careful explanation in this paragraph. When the energy band is parabolic, the current operator is proportional to the kinetic momentum operator: 
$\langle\vec{\mathbf{j}}(t)\rangle=-e\langle\vec{\mathbf{P}}(t)\rangle/m - ne^{2}\vec{\mathbf{A}}(t)/m$, where $\vec{\mathbf{P}}$ is the canonical momentum per unit volume. Since $\vec{\mathbf{P}}$ commutes with the Hamiltonian under uniform perturbation, its average value remains zero all the time. Thus the linear response is trivial and we got $\sigma(\omega)=ie^{2}n/m(\omega + i0^{+})$. We can see that there is only a delta function in the real part of the conductivity, and this derivation holds regardless of whether the ground state is a symmetry-breaking state or not.

\begin{figure}
\begin{center}
\includegraphics[width=3in]{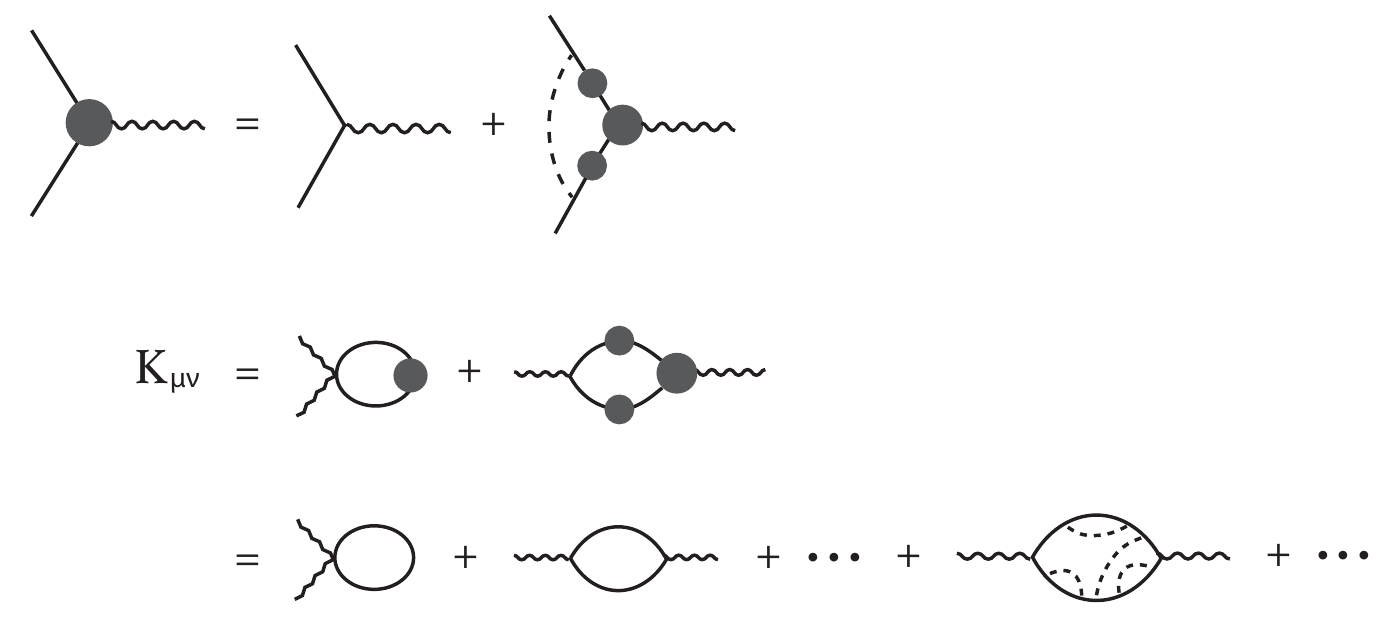}
\caption{The self-consistent vertex correction and the diagrams included in the corrected electromagnetic response function $K_{\mu\nu}$ (defined as $j_{\mu}=K_{\mu\nu}A_{\nu}$). The solid line represents the Nambu spinor, the dashed line represents the electron-electron interaction and the curly line represents the electromagnetic field. The second diagram on the first line of $K_{\mu\nu}$ is the paramagnetic response $P_{\mu\nu}$.}%
\label{fig:vertex}
\end{center}
\end{figure}

The inconsistencies (1) and (2) can be solved by a well-known technique in QED, first introduced to superconductors by Nambu to restore the gauge invariance in the BCS formalism \cite{peskin1995introduction,nambu1960quasi,schrieffer1964theory}.
The key observation is that, whenever an electron-photon vertex appears in a chain of electron lines, we can always form a `gauge-invariant subgroup' of diagrams by considering all different places to insert the corresponding photon line on this chain. And the Ward-Takahashi identity is automatically preserved if we sum over all diagrams in this subgroup. 
As discussed in the previous section, the mean field Green's function contains all diagrams without crossing. Following the diagrammatic technique, if we plug the mean field Green's function into the bubble diagram, we are forced to include all corrections to the bubble diagram without crossing. This can be done by introducing a corrected electron-photon vertex, as shown in figure \ref{fig:vertex}. Those diagrams containing a 2-electron-2-photon vertex correspond to the average value of the diamagnetic current, which does not contribute to the imaginary part of the response function (real part of the conductivity) at any finite frequency, so we focus on the paramagnetic part of the response function (defined as $j^{P}_{\mu}=P_{\mu\nu}A_{\nu}$):
\begin{eqnarray}
P_{\mu\nu}=-i\int\frac{d^{3}p}{(2\pi)^{3}}\text{Tr}[\gamma_{\mu}(p,p')G_{p'}\Gamma_{\nu}(p',p)G_{p}]
\end{eqnarray}
where $\gamma_{\mu}(p,p')$ ($\Gamma_{\mu}(p,p')$) is the bare (corrected) vertex of the 2-electron-1-photon interaction. $\Gamma_{\mu}(p,p')$ is given by a self-consistent equation as depicted in figure \ref{fig:vertex}: 
\begin{eqnarray}
\Gamma_{\mu}(p',p)&=&\gamma_{\mu}(p',p)+\nonumber\\
i\int\frac{d^{3}k}{(2\pi)^{3}}&\lambda_{k}&\tau_{3}G(p'\!-\!k)\Gamma_{\mu}(p'\!-\!k,p\!-\!k)G(p\!-\!k)\tau_{3}
\label{eq: vertex correction}
\end{eqnarray}
We are interested in the case $\vec{\mathbf{p}}=\vec{\mathbf{p}}'$, and we have $\gamma_{\mu}([p_{0}+\omega,\vec{\mathbf{p}}],[p_{0}, \vec{\mathbf{p}}])=\gamma_{\mu}(\vec{\mathbf{p}})$ as shown in equation \ref{eq:paramagnetic current} and \ref{eq:bare vertex}.

Equation \ref{eq: vertex correction} can be solved analytically when the four-Fermion interaction has no momentum dependence near the Fermi surface. If we further assume the pairing gap $\Delta$ is much smaller than the band width, the self-consistent vertex acquires a simple form:
\begin{eqnarray}
\label{eq:corrected vertex}
&\vec{\mathbf{\Gamma}}&=-e(\vec{\mathbf{v}}_{1}(\vec{\mathbf{p}})\mathbb{1}\! +\! \vec{\mathbf{v}}_{2}(\vec{\mathbf{p}})\tau_{3} + 2i\Delta I(\vec{\mathbf{v}}_{2})\tau_{2}/\omega I(1)),\\
&I&(f)\equiv \int_{\text{paired}}\frac{d^{2}\vec{\mathbf{p}}}{(2\pi)^{2}}\frac{f(p)}{\delta_{p}(\omega-2\delta_{p})(\omega+2\delta_{p})}
\label{eq:integral}
\end{eqnarray}
where $I(f)$ is a linear functional defined by the integral which appears repeatedly in the remaining part of the paper. Finally the corrected optical conductivity is given by
\begin{eqnarray}
\text{Re}\,&\sigma_{ij}&(\omega>0)= -\text{Im}\,P_{ij}(\omega>0)/\omega\\
&=&-\frac{4e^{2}\Delta^{2}}{\hbar\omega}\text{Im}\left[I(v_{2i}v_{2j})-I(v_{2i})I(v_{2j})/I(1)\right]
\label{eq: final results}
\end{eqnarray}
Note that we have omitted the infinitesimal imaginary part of $\omega$ in the integral \ref{eq:integral} since the pole structure in retarded response functions is different from that in path integrals, and $\omega$ should always be replaced by $\omega + i0^{+}$ for retarded response. When $\omega>0$, the imaginary part of the integral is given by
\begin{equation}
\text{Im}\,I(f)=-\pi\int_{\text{paired}}\frac{d^{2}\vec{\mathbf{p}}}{(2\pi)^{2}}\frac{f(p)}{4\delta_{p}^{2}}\delta(\omega-2\delta_{p})
\end{equation}
which is proportional to the joint density of states (JDOS) in the paired region. We found that the first term in equation \ref{eq: final results} is nothing but the bare result given by the `mean-field-version' of equation \ref{eq: bare result}, while the second term is given by the vertex correction. As discussed before, only those points in the `paired region' of the B.Z., where the frequency matches the band splitting, contribute to the real part of the optical conductivity. For a given $\omega$, these points lie on arcs in the B.Z.

Another important ingredient in equation \ref{eq: final results} is $\vec{\mathbf{v}}_{2}$. Recall that $\vec{\mathbf{v}}_{2}$ is defined by equation \ref{eq:current} and \ref{eq:current,Nambu spinor}. In the case of FFLO pairing, when the pairing momentum is much smaller than the Fermi momentum, we have
\begin{equation}
v_{2i}(\vec{\mathbf{p}})=(\mathbf{m}_{p}^{-1})_{ij}Q_{j}/2 + O(Q^{2})
\end{equation}
As discussed above, gauge invariance is guaranteed in this formalism. Furthermore, we found that the problem regarding Galilean symmetry is automatically solved: if the band is parabolic, $\vec{\mathbf{v}}_{2}=\vec{\mathbf{Q}}/2m=\text{const.}$, hence $v_{2i}$ and $v_{2j}$ can be dragged out of the integral in equation \ref{eq: final results}, and the vertex correction cancels the bare result. We refer the readers to the appendix for more details on the Ward-Takahashi identity, the vertex correction and the final result for optical conductivity.

\section{Results for tight-binding bands}
\begin{figure}
\begin{center}
\includegraphics[width=3in]{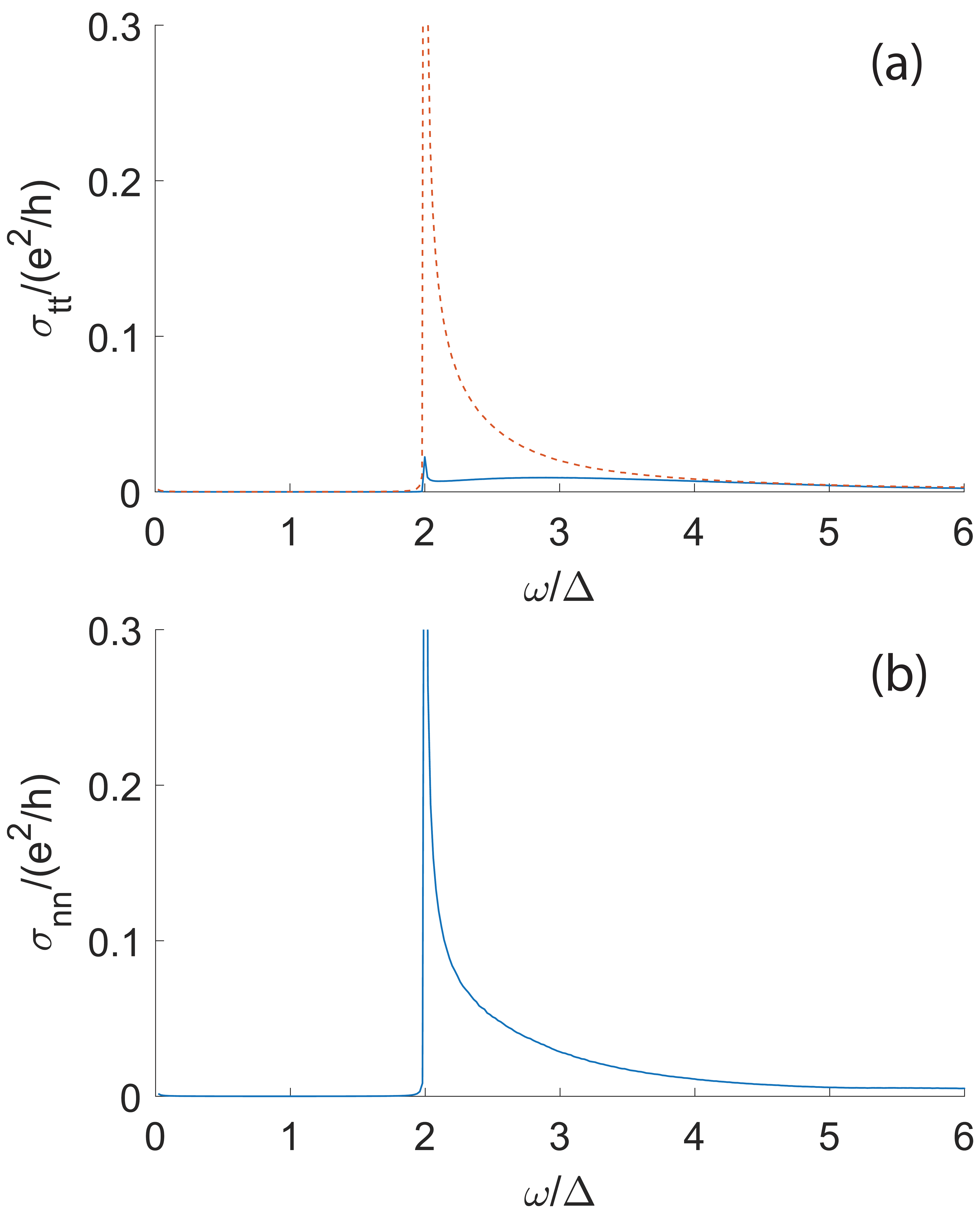}
\caption{Optical conductivity of the FFLO state calculated for tight-binding bands on a 2-dimensional square lattice, $t_{2}/t_{1}=0.35$. The spin splitting is set to be $0.4t_{1}$, which is about four percent of the band width, and the pairing momentum is $(0.1/a, 0.1/a)$. (a) Conductivity in the direction of the pairing momentum. The dashed orange line is the bare result and the blue line is the corrected result. (b) Conductivity in the perpendicular direction. The vertex correction is identically zero in this direction by symmetry.}%
\label{fig:result}
\end{center}
\end{figure}
We have calculated the optical conductivity of FFLO states explicitly for tight-binding bands with NN hoping $t_{1}$ and NNN hoping $t_{2}$ on a square lattice. The result shown in figure \ref{fig:result} is for $t_{2}/t_{1}=0.35$, spin splitting $0.4t_{1}$, at half-filling. The pairing momentum is $(0.1/a,0.1/a)$, where $a$ is the lattice constant. AC conductivity shows up above $2\Delta$ and there is a divergent peak right at $2\Delta$ due to the corresponding divergence in the JDOS. As mentioned in the previous section, for a given $\omega$, only the arcs in the B.Z. satisfying the frequency-matching condition contribute to AC absorption. When $\delta\omega\equiv\omega-2\Delta\simeq 0$, the frequency-matching condition $\omega=2\delta_{p}$ gives $\bar{\epsilon}_{p}=\sqrt{\omega^{2}/4-\Delta^{2}}\propto\sqrt{\delta\omega}$, then the JDOS is $N(0)d\bar{\epsilon}_{p}/d\omega\propto 1/\sqrt{\delta\omega}$, where $N(0)$ is the density of states (DOS) of the normal metal. Hence the $1/\sqrt{\delta\omega}$ divergence in the optical conductivity at $2\Delta$. This divergence has the same form of the divergence in the DOS and JDOS of s wave BCS superconductors, but the real part of the AC conductivity is identically zero in BCS superconductors as explained in the introduction.

The effects of the vertex correction on divergent peaks depend on the type of divergence as well as the details of the band structure, and can be dramatically different in different situations. If there is a single singularity of the JDOS on the frequency-matching arc giving the dominant contribution, we can replace $\vec{\mathbf{v}}_{2}$ by its value at the singularity, and it is clear from equation \ref{eq: final results} that the vertex correction completely cancels the divergence in the bare result. However, the divergence at $2\Delta$ is due to the whole arc in the paired region satisfying $\bar{\epsilon}_{p}\simeq 0$, and it remains divergent after the vertex correction. The ratio between the corrected result (shown as blue line in figure \ref{fig:result}) and the bare result (dashed orange line) depends on the variance of $\vec{\mathbf{v}}_{2}$ on the frequency-matching arc.  We found that in the current example, the divergence in the conductivity along the pairing momentum $\sigma_{tt}$ is strongly suppressed by the vertex correction, whereas there is no vertex correction at all in the perpendicular direction since the perpendicular component of $\vec{\mathbf{v}}_{2}$ is odd under the reflection over $(\pi,\pi)$.

\section{Discussion}

We have shown that there is nonzero AC absorption from PDW if we break Galilean symmetry explicitly in the electronic Hamiltonian (which is usually the case in solids). When the pairing momentum $Q$ is much smaller than the Fermi momentum $p_{F}$ and the pairing gap $\Delta$ is much smaller than the band width $W$, the AC conductivity is proportional to $(Q/p_{F})^{2}W/\Delta$. Vertex correction plays an important role in this AC absorption, and dramatically changes the behavior of the optical conductivity in the direction of the pairing momentum. 

This nonzero absorption could be used as an experimental evidence for PDW. Furthermore, the various features discussed in the previous section can help determine the pairing gap and the direction of the pairing momentum in experiments. We have focused on the case with only one pairing momentum in the present paper, and we have ignored the momentum dependence of the pairing gap near Fermi surface in the explicit calculation. The results for more general PDW should be similar, but we would like to discuss some possible differences in this paragraph. (1) A weak momentum dependence of the pairing gap introduces a cutoff to the $1/\sqrt{\delta\omega}$ divergence at $\omega=2\min[\Delta_{p}]$, whereas a strong momentum dependence completely destroys the $1/\sqrt{\delta\omega}$ behavior and leaves only a finite jump. (2) When the PDW state has more than one pairing momenta, one or more CDW will be generated by the interference, and there will be nonzero absorption below the `pairing gap' $2\min[\Delta_{p}]$. The magnitude of this `in gap' absorption increases with the magnitude of the CDW. (3) We have not discussed the effect of impurities so far. Since there is a finite density of states left at Fermi level, there will be a Drude peak coexisting with the absorption we discussed when the inverse of the mean free time of electrons is smaller than the pairing gap. Whereas in the opposite limit, even BCS superconductors have nonzero optical absorption above the gap \cite{mahan2013many} and there is no sharp feature for PDW.

PAL acknowledges support by NSF under DMR-1522575.

\bibliographystyle{apsrev4-1}
\bibliography{OCPDWref}

\onecolumngrid
\section{Appendix}

We present the derivation of equation \ref{eq:corrected vertex} and \ref{eq: final results} in this appendix. For simplicity, we define $\tilde{p}_{0}\equiv p_{0}-\epsilon'_{p}$. The Green's function given by equation \ref{eq:mean field Green} can then be written as:
\begin{equation}
G(p)=\frac{1}{\tilde{p}_{0}-\bar{\epsilon}_{p}\tau_{3}-\Delta_{p}\tau_{1}+isgn(p_{0})0^{+}}=\frac{\tilde{p}_{0}+\bar{\epsilon}_{p}\tau_{3}+\Delta_{p}\tau_{1}}{(\tilde{p}_{0}+isgn(p_{0})0^{+})^{2}-\delta_{p}^{2}}
\end{equation}
where we have neglected the diagonal self-energy correction since it is not important for our purpose. We are free to choose the `direction' of the pairing term in the $\tau_{1}-\tau_{2}$ plan since they are related by gauge symmetry. The temporal component of the self-consistent vertex $\Gamma_{t}$ in the limit $|\vec{\mathbf{q}}|\to 0$ ($q$ is the momentum of the external field) is determined directly by the Ward-Takahashi identity 
\begin{equation}
q_{\mu}\Gamma_{\mu}(p+q,p)=-e\tau_{3}G^{-1}(p)+eG^{-1}(p+q)\tau_{3}
\end{equation}
Where $q_{\mu}\Gamma_{\mu}$ is a shorthand for $\vec{\mathbf{q}}\cdot\vec{\mathbf{\Gamma}}-\omega\Gamma_{t}$. Note that there are additional $\tau_{3}$'s compared to the standard Ward-Takahashi identity in QED since the two components of the Nambu spinor carry opposite charges. If we assume the spatial components of $\Gamma$ does not diverge in the limit $|\vec{\mathbf{q}}|\to 0$, which can be verified latter, only the temporal component of $\Gamma$ contribute the left hand side, and we have
\begin{equation}
\Gamma_{t}([p_{0}+\omega,\vec{\mathbf{p}}],[p_{0}, \vec{\mathbf{p}}])=-(-e\tau_{3}G^{-1}(p)+eG^{-1}(p+q)\tau_{3})/\omega=-e(\tau_{3}+2i\Delta_{p}\tau_{2}/\omega)
\end{equation}
On the other hand, the spatial components of $\Gamma$ take some calculation, and they acquire a simple form only when the four-Fermion interaction has no momentum dependence near the Fermi surface. In this case $\lambda_{k}$ can be treated as a constant, and the self-consistent equation (equation \ref{eq:mean field Green}) shows that $\Delta_{p}$ is also a constant near the Fermi surface. Plugging the mean field Green's function in equation \ref{eq: vertex correction}, and shifting the momentum of the integration, we have
\begin{equation}
\Gamma_{\mu}([p_{0}+\omega,\vec{\mathbf{p}}],[p_{0}, \vec{\mathbf{p}}])=\gamma_{\mu}(\vec{\mathbf{p}})+i\lambda\int\frac{d^{3}p}{(2\pi)^{3}}\frac{\tau_{3}(\tilde{p}_{0}+\omega+\bar{\epsilon}_{p}\tau_{3}+\Delta\tau_{1})\Gamma_{\mu}([p_{0}+\omega,\vec{\mathbf{p}}],[p_{0}, \vec{\mathbf{p}}])(\tilde{p}_{0}+\bar{\epsilon}_{p}\tau_{3}+\Delta\tau_{1})\tau_{3}}{((\tilde{p}_{0}+\omega+isgn(p_{0}+\omega)0^{+})^{2}-\delta_{p}^{2})((\tilde{p}_{0}+isgn(p_{0})0^{+})^{2}-\delta_{p}^{2})}
\label{extremely long}
\end{equation}
It is clear from the equation above that the vertex correction has no p dependence, this is of course only true when we ignore the momentum dependence of the four-Fermion interaction. In this case, we can write the self-consistent vertex as
\begin{equation}
\Gamma_{\mu}([p_{0}+\omega,\vec{\mathbf{p}}],[p_{0}, \vec{\mathbf{p}}])=\gamma_{\mu}(\vec{\mathbf{p}})-e\Gamma_{\mu}^{0}\mathbb{1}-e\sum_{i=1}^{3}\Gamma_{\mu}^{i}\tau_{i}
\label{eq:Gammai}
\end{equation}
where $\Gamma^{0}$ and $\Gamma^{i}$ are functions of $\omega$, and $\gamma_{\mu}(\vec{\mathbf{p}})$ is given by equation \ref{eq:paramagnetic current} and \ref{eq:bare vertex}. The next step is to plug equation \ref{eq:Gammai} into equation \ref{extremely long}, compute the matrix multiplication in the numerator, carry out the integral of $p_{0}$ using the residue theorem and solve $\Gamma^{0}$ and $\Gamma^{i}$. Note that there are 4 poles of $p_{0}$ in the complex plane, whose imaginary parts depend on the spatial momentum $\vec{\mathbf{p}}$. If $\vec{\mathbf{p}}$ lies in the `unpaired region', the two eigenenergies $E_{p}^{\pm}$ are of the same sign, so the four poles locate at the same side of the real axis. Then we know the integral must be zero since we can complete the contour on the other side including none of the residues. This observation confirms our statement that only the `paired region' in the B.Z. contribute to the optical conductivity. After all these laborious calculation, we arrive at the self-consistent equation for $\vec{\mathbf{\Gamma}}^{0}$ and $\vec{\mathbf{\Gamma}}^{i}$ (the spatial components of $\Gamma^{0}$ and $\Gamma^{i}$). We showed that, by direct calculation, the integral in equation \ref{extremely long} has no identity component, thus $\vec{\mathbf{\Gamma}}^{0}=0$. On the other hand, $\vec{\mathbf{\Gamma}}^{i}$ satisfies
\begin{eqnarray}
\left(\begin{array}{c}
 \vec{\mathbf{\Gamma}}^{1}\\ \vec{\mathbf{\Gamma}}^{2}\\ \vec{\mathbf{\Gamma}}^{3}
\end{array}\right)&=& \lambda
\left(\begin{array}{ccc}
2I(\bar{\epsilon}_{p}^{2}) & -i\omega I(\bar{\epsilon}_{p}) & -2\Delta I(\bar{\epsilon}_{p})\\
i\omega I(\bar{\epsilon}_{p}) & 2I(\delta_{p}^{2}) & -i\omega\Delta I(1)\\
2\Delta I(\bar{\epsilon}_{p}) & -i\omega\Delta I(1) & -2\Delta^{2}I(1)
\end{array}\right)
\left(\begin{array}{c}
 \vec{\mathbf{\Gamma}}^{1}\\ \vec{\mathbf{\Gamma}}^{2}\\ \vec{\mathbf{\Gamma}}^{3}
\end{array}\right) + \lambda
\left(\begin{array}{c}
 -2\Delta I(\bar{\epsilon}_{p}\vec{\mathbf{v}}_{2})\\
 -i\omega\Delta I(\vec{\mathbf{v}}_{2})\\ 
 -2\Delta^{2}I(\vec{\mathbf{v}}_{2})
\end{array}\right) \label{eq:matrix}\\
\text{where}\ I(f(\vec{\mathbf{p}}))&\equiv& \int_{\text{paired}}\frac{d^{2}\vec{\mathbf{p}}}{(2\pi)^{2}}\frac{f(p)}{\delta_{p}(\omega-2\delta_{p}+isgn(\omega)0^{+})(\omega+2\delta_{p}+isgn(\omega)0^{+})}
\label{eq:If, path integral}
\end{eqnarray}
If we further assume the pairing gap $\Delta$ and the frequency $\omega$ is much smaller than the band width, only a thin shell near $\bar{\epsilon}_{p}=0$ contribute to the integral. In this limit $I(\bar{\epsilon}_{p})\sim 0,\ I(\bar{\epsilon}_{p}\vec{\mathbf{v}}_{2})\sim 0$, so we have $\vec{\mathbf{\Gamma}}^{1}\sim 0$, $\vec{\mathbf{\Gamma}}^{2}$ and $\vec{\mathbf{\Gamma}}^{3}$ satisfies
\begin{eqnarray}
\left(\begin{array}{c}
 \vec{\mathbf{\Gamma}}^{2}\\ \vec{\mathbf{\Gamma}}^{3}
\end{array}\right)= \lambda
\left(\begin{array}{cc}
2I(\delta_{p}^{2}) & -i\omega\Delta I(1)\\
-i\omega\Delta I(1) & -2\Delta^{2}I(1)
\end{array}\right)
\left(\begin{array}{c}
 \vec{\mathbf{\Gamma}}^{2}\\ \vec{\mathbf{\Gamma}}^{3}
\end{array}\right) - \lambda I(\vec{\mathbf{v}}_{2})
\left(\begin{array}{c}
 i\omega\Delta \\ 
 2\Delta^{2}
\end{array}\right)
\end{eqnarray}
In addition, the mean field gap equation gives us
\begin{equation}
4\lambda I(\delta_{p}^{2})-\lambda\omega^{2}I(1)=-\lambda I(\omega^{2}-4\delta_{p}^{2})=-2\lambda\int_{\text{paired}}\frac{d^{2}\vec{\mathbf{p}}}{(2\pi)^{2}}\frac{1}{2\sqrt{\bar{\epsilon}_{p}^{2} + \Delta^{2}}}=2
\end{equation}
Using this identity, we can easily find
\begin{eqnarray}
\left\{\begin{array}{l}
\vec{\mathbf{\Gamma}}^{2}=\frac{2i\Delta I(\vec{\mathbf{v}}_{2})}{\omega I(1)}\\
\vec{\mathbf{\Gamma}}^{3}=0
\end{array}\right.
\end{eqnarray}
So the corrected vertex is
\begin{equation}
\Gamma_{\mu}([p_{0}+\omega,\vec{\mathbf{p}}],[p_{0}, \vec{\mathbf{p}}])=-e[\tau_{3}+2i\Delta\tau_{2}/\omega,\ \vec{\mathbf{v}}_{1}(\vec{\mathbf{p}})\mathbb{1} + \vec{\mathbf{v}}_{2}(\vec{\mathbf{p}})\tau_{3} + 2i\Delta I(\vec{\mathbf{v}}_{2})\tau_{2}/\omega I(1)]
\end{equation}
We are now ready to calculate the paramagnetic response function $P_{\mu\nu}$. For simplicity, define
\begin{equation}
\langle f,h\rangle\equiv-i\int\frac{d^{3}p}{(2\pi)^{3}}Tr[f(p,p')G_{p'}h(p',p)G_{p}]
\end{equation}
Then we have
\begin{eqnarray}
P_{ij}&=&\langle \gamma_{i},\Gamma_{j}\rangle\\
&=& e^{2}\langle v_{1i}(\vec{\mathbf{p}})\mathbb{1} + v_{2i}(\vec{\mathbf{p}})\tau_{3} ,v_{1j}(\vec{\mathbf{p}})\mathbb{1} + v_{2j}(\vec{\mathbf{p}})\tau_{3} + 2i\Delta I(v_{2j})\tau_{2}/\omega I(1)\rangle\\
&=& e^{2}\langle v_{2i}(\vec{\mathbf{p}})\tau_{3},v_{2j}(\vec{\mathbf{p}})\tau_{3}\rangle + (2i\Delta I(v_{2j})/\omega I(1))e^{2}\langle v_{2i}(\vec{\mathbf{p}})\tau_{3}, \tau_{2}\rangle
\end{eqnarray}
where we have used the fact that the identity component of the vertex does not contribute to the integral, which can be verified explicitly. Integrating out $p_{0}$ we have
\begin{equation}
P_{ij}=4e^{2}\Delta^{2}\left[I(v_{2i}v_{2j})-I(v_{2i})I(v_{2j})/I(1)\right]
\end{equation}
This result leads to the result for optical conductivity in equation \ref{eq: final results}. We would like to remind the readers again that equation \ref{eq: final results} holds only for $\omega>0$ if we define the integral $I(f(\vec{\mathbf{p}}))$ as in equation \ref{eq:If, path integral}, this is due to the difference between path integral and retarded response. It holds for both positive and negative $\omega$ if we replace the infinitesimal imaginary part $isgn(\omega)0^{+}$ in the integral $I(f(\vec{\mathbf{p}}))$ by $i0^{+}$.
\end{document}